\documentclass[final]{svjour2}
\usepackage{graphicx}
\usepackage{rotating}
\usepackage{amssymb}
\usepackage{mathptmx}
\usepackage[numbers]{natbib}
\makeatletter
\journalname{Journal of Low Temperature Physics}

\bibpunct{}{}{,}{s}{}{,}

\begin{document}

\newcommand{\hdblarrow}{H\makebox[0.9ex][l]{$\downdownarrows$}-}
\title{Evidence of Novel Quasiparticles in a Strongly Interacting Two-Dimensional Electron System: Giant Thermopower and Metallic Behaviour}

\author{V. Narayan$^1$ \and M. Pepper$^2$ \and J. Griffths$^1$ \and H. Beere$^1$ \and F. Sfigakis$^1$ \and G. Jones$^1$ \and D. Ritchie$^1$ \and
A. Ghosh$^3$}

\institute{1:Cavendish Laboratory, University of Cambridge, UK\\
Tel.:00441223337469\\ Fax:00441223337271\\
\email{vn237@cam.ac.uk}
\\2:Department of Electronic and Electrical Engineering, University College London, UK\\ 3: Department of Physics, Indian Institute of Science}

\date{07.08.2012}

\maketitle


\begin{abstract}
We report thermopower ($S$) and electrical resistivity ($\rho_{2DES}$) measurements in low-density (10$^{14}$~m$^{-2}$), mesoscopic two-dimensional electron systems (2DESs) in GaAs/AlGaAs heterostructures at sub-Kelvin temperatures. We observe at temperatures $\lesssim$ 0.7~K a linearly growing $S$ as a function of temperature indicating metal-like behaviour. Interestingly this metallicity is not Drude-like, showing several unusual characteristics: i) the magnitude of $S$ exceeds the Mott prediction valid for non-interacting metallic 2DESs at similar carrier densities by over two orders of magnitude; and ii) $\rho_{2DES}$ in this regime is two orders of magnitude greater than the quantum of resistance $h/e^2$ and shows very little temperature-dependence. We provide evidence suggesting that these observations arise due to the formation of novel quasiparticles in the 2DES that are \textit{not} electron-like. Finally, $\rho_{2DES}$ and $S$ show an intriguing decoupling in their density-dependence, the latter showing striking oscillations and even sign changes that are completely absent in the resistivity.

\end{abstract}

\section{Introduction}

It has been previously observed in low-density 2DESs of mesoscopic dimensions that $\rho_{2DES}$ saturates or even \textit{decreases} upon lowering the temperature $T$ below $\approx$ 1~K~\cite{Matthias} suggesting a destabilisation of the insulating phase. This is especially remarkable in light of the fact that at these low values of density ($n_s \sim$ 10$^{14}$~m$^{-2}$) $\rho_{2DES} \gg h/e^2$ where it is expected that the 2DES be strongly localised and the transport acivated. The density-dependence of $\rho_{2DES}$ in this regime has been observed to obey a Kosterlitz-Thouless law suggesting that the 2DES is undergoing a disorder-to-order transition as $n_s$ is decreased~\cite{Koushik}. It is well-known that the thermopower is more sensitive to inter-electron interactions than the electrical resistivity~\cite{ChaikinBurns} and, as explained in the following, contains complementary information. The Mott formula~\cite{Mott} relates the diffusion thermopower $S_{MOTT}$ and the electrical conductivity $\sigma$ of a system as:

\begin{equation}
\label{Mottformula}
S_{MOTT} = \frac{\pi^2 k_BT}{3q}\left(\frac{\partial \ln \sigma}{\partial E}\right)_{E = \mu}
\end{equation}

Here $k_B$ is the Boltzmann constant, $q$ is the charge of the carriers, $E$ is the energy and $\mu$ is the chemical potential os the system. Thus we see that $S_d$ contains complementary information to $\sigma$ being, as it is, sensitive to its energy-derivative. Motivated by this, we have recently performed thermopower measurements on low-density, mesoscopic 2DESs to investigate many-body effects in them~\cite{Narayan}. In the present manuscript we first briefly review our results and then present a picture that self-consistently explains the observations in refs.~\cite{Matthias, Koushik, Narayan}.

\section{Experimental details}

\begin{figure}
\begin{center}
\includegraphics[%
  width=1.0\linewidth,
  keepaspectratio]{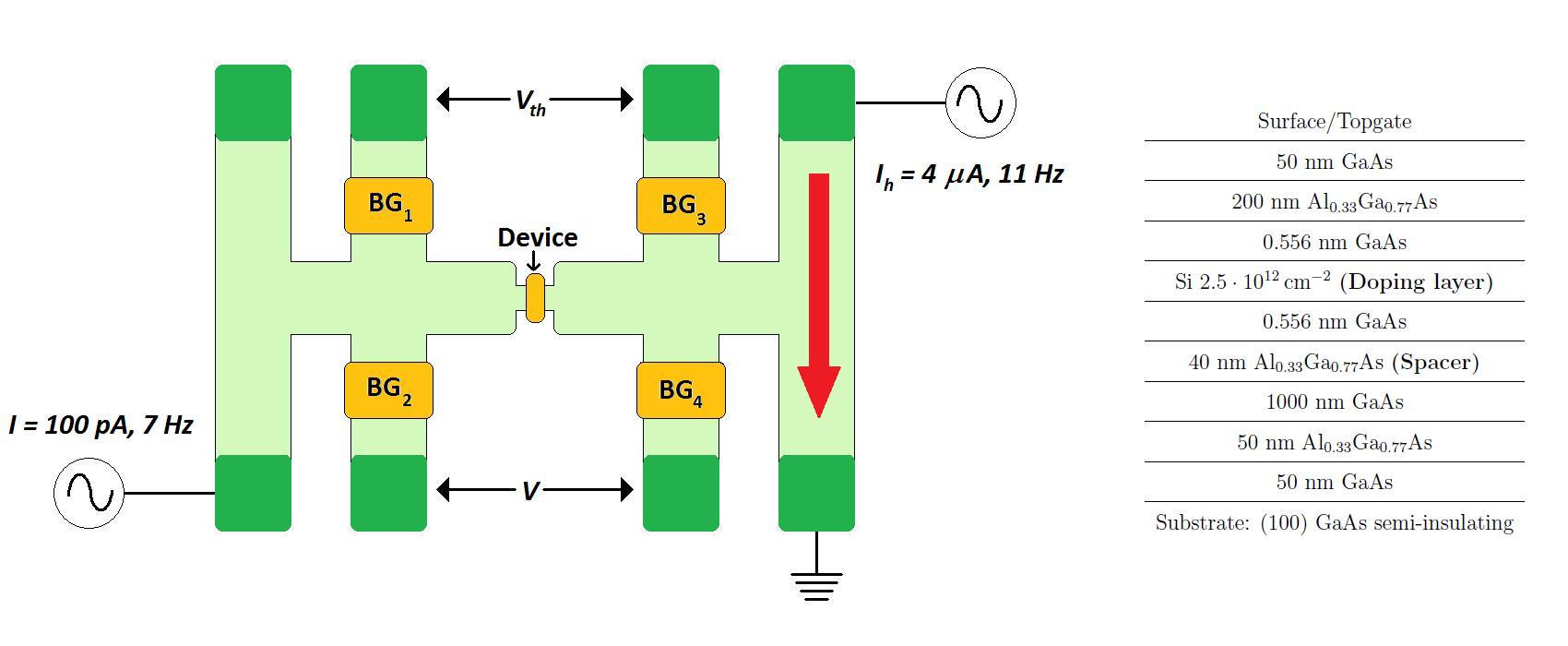}
\end{center}
\caption{(Colour online) The left panel shows a schematic of the device layout and measurement setup. The device is in a Hall-Bar layout with ohmic contacts (marked as `X') and metallic gates (shown in yellow). BG$_1$ -- BG$_4$ are bar-gates used for the electron thermometry and the mesoscopic 2DES is labeled ``Device''. As shown, $\rho_{2DES}$ is measured in a constant-current ($I$) setup and the thermovoltage $V_{th}$ is detected in response to the heating due to $I_h$. The right panel shows the wafer structure used in the present experiments.}
\label{Device}
\end{figure}

A schematic represenation of the device and measurement setup are given in figure~\ref{Device}. Our measurement devices are fabricated in Si-doped GaAs/AlGaAs heterostructures in which the dopants are confined to a monolayer 40~nm above the 2DES. The as-grown mobility of the wafer is 220~m$^2$/Vs at a carrier density $=$~2.2 $\times$ 10$^{15}$~m$^{-2}$. We used optical and electron-beam lithography to define the mesoscopic 2DES (L$\times$W = 1-2~$\mu$m~$\times$~8~$\mu$m). $\rho_{2DES}$ was measured in a 4-probe setup using a lock-in amplifier at 7~Hz. To measure $S$ we imposed a temperature gradient along its length using a large heating current $I_h$~= 4-5~$\mu$A at $f_h$~=11~Hz and detected the thermovoltage $V_{th}$ using a lock-in amplifier at 2$f_h$. To measure the temperature difference $\Delta T$ across the device, we adapted a method following refs.~\cite{Appleyard, Chickering} wherein we measured the thermovoltage due to $\Delta T$ between pairs of large bar gate-defined regions ($\approx$~10~$\mu$m in linear dimension) biased such that the 2DESs underneath them are in the non-interacting Drude-metal-like regime. Under such circumstances, the thermopower is given by:

\begin{equation}
\label{DrudeMottformula}
S_d = -\frac{\pi k_B^2Tm}{3|e|\hbar^2}\frac{1 + \alpha}{n}
\end{equation}

Here $m$ is the effective mass of the carriers, $n$ is the density of carriers and $\alpha \equiv (n/\tau)(d\tau/dn)$ where $\tau$ is the momentum scattering time. Further details regarding the measurement process can be found in ref.~\cite{Narayan}.

\section{Results}

\begin{figure}
\begin{center}
\includegraphics[%
  width=1.0\linewidth,
  keepaspectratio]{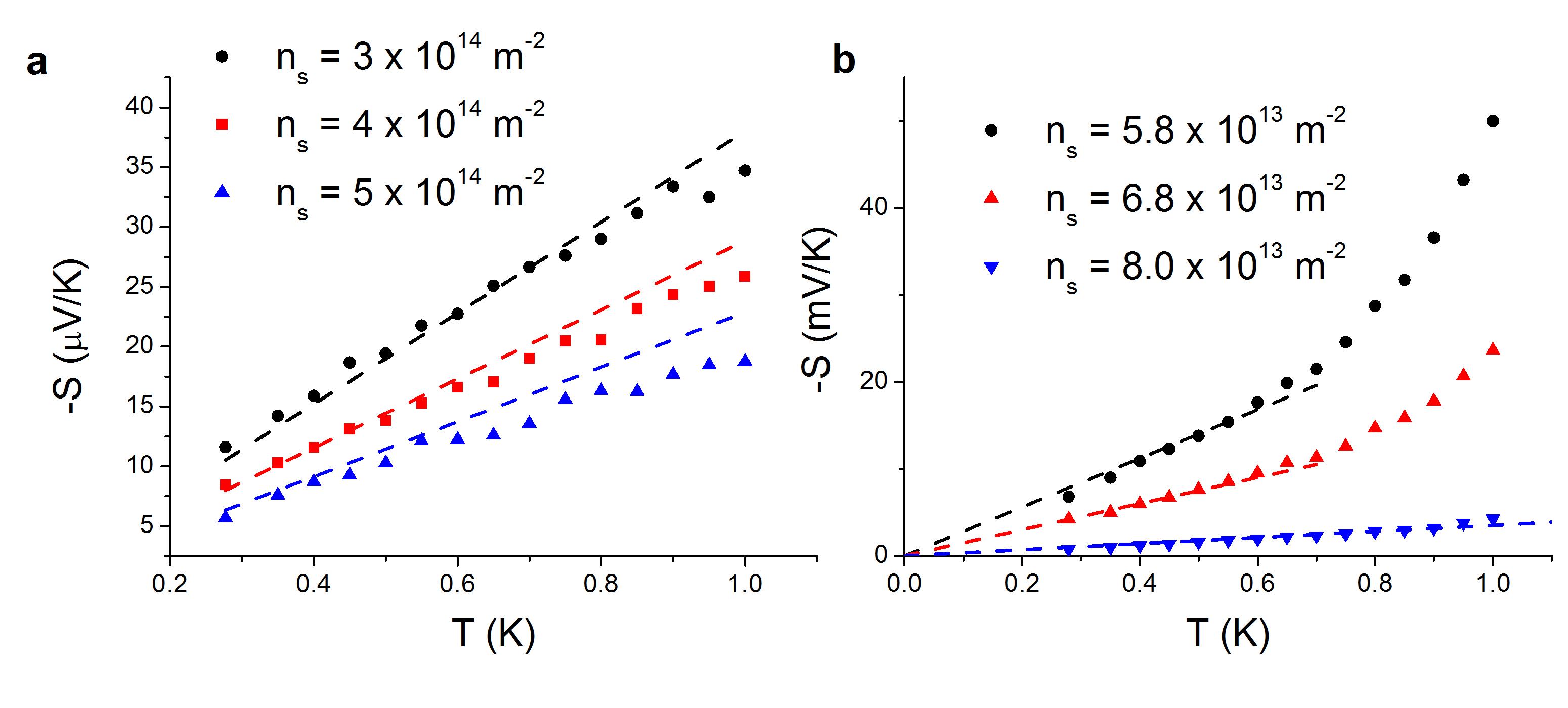}
\end{center}
\caption{(Colour online) $T$-dependence of $S$. As is seen, $S$ is linear against $T$ in both the $\rho_{2DES} < h/e^2$ and $\rho_{2DES} \gg h/e^2$ regimes. a) The broken lines represent the Mott prediction $S_d$ (equation~\ref{DrudeMottformula}). When $\rho_{2DES} < h/e^2$, we see that the Mott relation holds quantitatively. b) The broken lines represent linear fits to the low-$T$ data. When $\rho_{2DES} \gg h/e^2$ while the linear behaviour is still seen below $\lesssim$ 0.7~K, $S$ is more than 2 orders of magnitude larger than $S_d$ (not visible on the scale of the figure).}
\label{TempDependence}
\end{figure}

In figure~\ref{TempDependence} we show the behavior of $S$ as a function of $T$. The two panels contrast the behaviour in the two regimes where $\rho_{2DES} \ll h/e^2$ and $\rho_{2DES} \gg h/e^2$. It is seen that in the former, the observed $S$ agrees closely with $S_d$ (equation~\ref{DrudeMottformula}) while in the latter, while $S(T)$ is still linear for $T \lesssim$ 0.7~K, $S$ is 2 orders of magnitude larger than $S_d$. Above 0.7~K $S$ grows at a much more rapid rate attaining exceedingly large values of $\approx 50$~mV/K at 1~K. Super-linear growths in $S$ are often associated with phonon-drag~\cite{FanielPRL2005, LyoPRB1988}, however, no similar increased growth rate is observed at higher $n_s$ (see figure~\ref{TempDependence}a), leaving open the possibility that the observed phonomenon is diffusive in nature. The agreement between $S$ and $S_d$ at high $n_s$ serves as a good control experiment that validates the $\Delta T$ measurement and eliminates the possibility of artifacts arising due to sample parameters and device geometry.

\begin{figure}[b]
\begin{center}
\includegraphics[%
  width=1.0\linewidth,
  keepaspectratio]{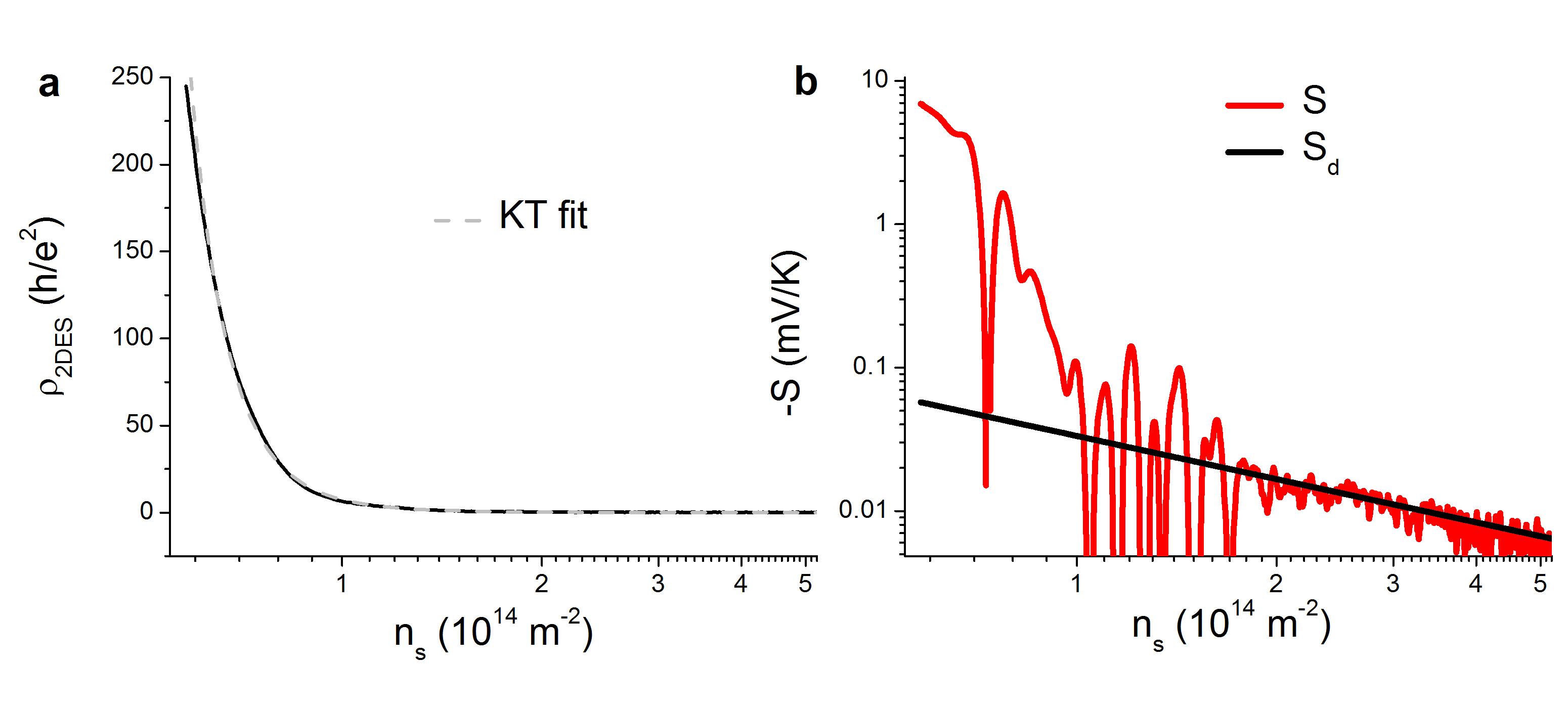}
\end{center}
\caption{(Colour online) $n_s$-dependence of $\rho_{2DES}$ and $S$. a) $\rho_{2DES}$ is seen to grow monotonically as $n_s$ is decreased. The broken line shows a fit to the Kosterlitz-Thouless form (see text for details). b) $S$ is seen to track $S_d$ (equation~\ref{DrudeMottformula} closely for densities above $\approx$ 1.7~$\times$ 10$^{14}$~m$^{-2}$ below which it oscillates, even changes sign and exceed $S_d$ by over two orders of magnitude. $\rho_{2DES}$ and and $S$ are measured simultaneously in the data presented here.}
\label{DensityDependence}
\end{figure}

In figures~\ref{DensityDependence}a and \ref{DensityDependence}b we plot the $n_s$-dependence of $\rho_{2DES}$ and $S$ at $T$ = 0.3~K, respectively. There are several points worth noticing: 1) $\rho_{2DES}$ grows monotonically as a function of $n_s$ over the entire range. 2) In sharp contrast to this, $S$ is highly non-monotonic over the same $n_s$-range, frequently changing sign. 3) $S$ tracks $S_d$ closely until $\approx$ 2 $\times$ 10$^{14}$~m$^{-2}$ around which point the oscillations set in. 4) As $n_s$ is lowered further $S$ increases rapidly in magnitude, exceeding $S_d$ by 2 orders of magnitude at the lowest $n_s$.

The data presented shows a measured breakdown of the Mott formula, equation~\ref{Mottformula}. The Mott formula is based on semiclassical, Boltzmann-like transport and thus we can expect departures from it when interaction effects are strong as is the case in the present study. We first address the large magnitude of $S$. In order to do so, we inspect the $n_s$-dependence of $\rho_{2DES}$ and, as noted in refs.~\cite{Koushik, Narayan} we see that the dependence of $\rho_{2DES}$ on $n_s$ is well fit by a Kosterlitz-Thouless~\cite{KT} law,

\begin{equation}
\label{KT}
\rho_{2DES} = \rho_0 \exp{(A/\sqrt{n_s - n_{KT}})}
\end{equation}

\begin{figure}
\begin{center}
\includegraphics[%
  width=0.5\linewidth,
  keepaspectratio]{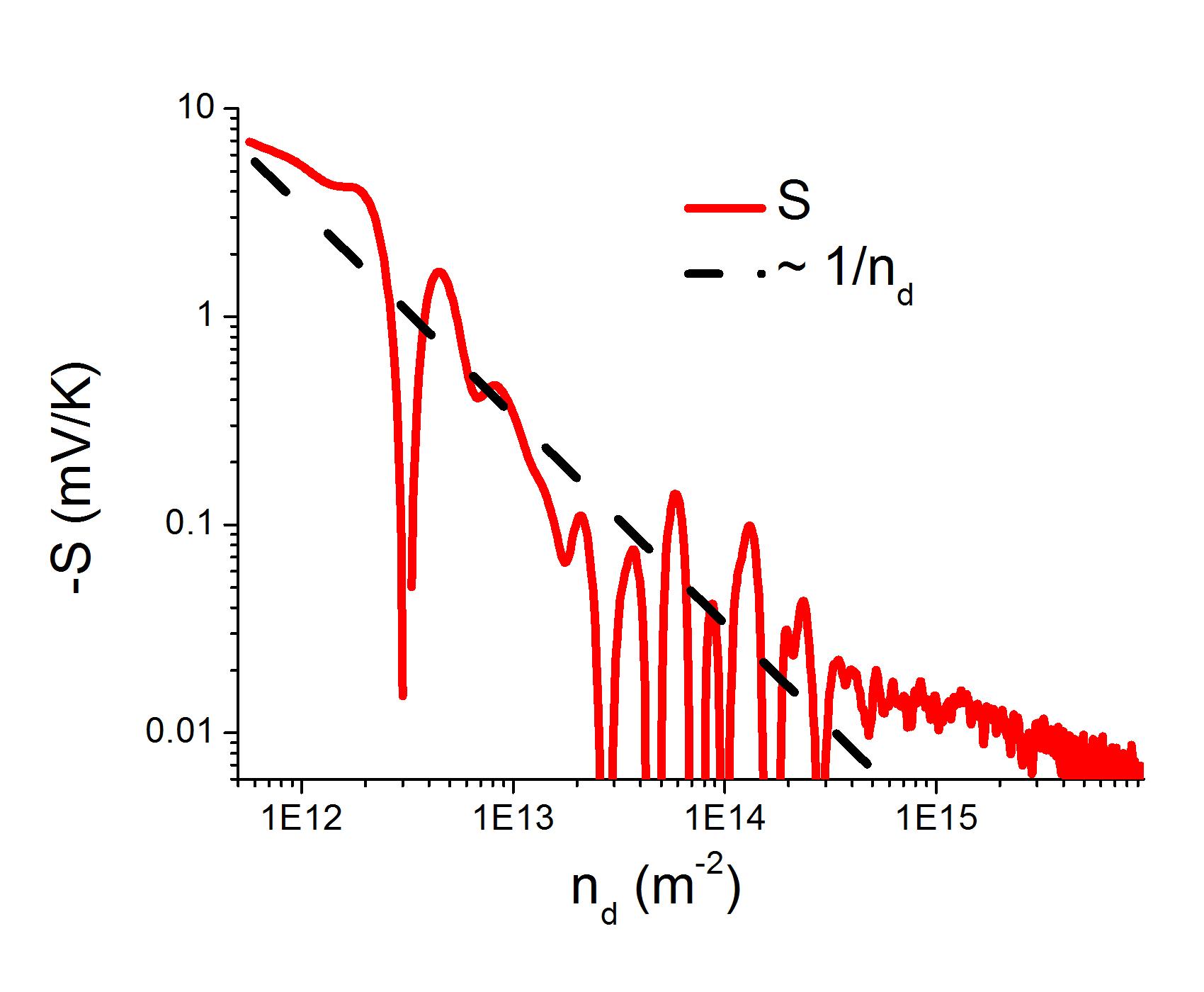}
\end{center}
\caption{(Colour online) Topological defect-mediated transport? The figure shows the dependence of $S$ on the number of topological defects $n_d$ (see text). Despite the strong oscillations, $S$ is seen to follow a clear $1/n_d$ trend. While the estimated value of $n_d$ is defined only to within an unknown constant, it is important to note that this has no bearing on the $1/n_d$ dependence.}
\label{SvsNd}
\end{figure}

Here $\rho_0$, $A$ and $n_{KT}$ are fit parameters, $n_{KT}$ being the transition density. As suggested in ref.~\cite{Koushik}, this implies that as $n_s$ is reduced towards $n_{KT}$ the 2DES approaches an ordered phase by the systematic elimination of topological defects and the topological defects rather than the bare electrons that mediate the electrical transport. While the exact nature of the ordered state cannot be ascertained using transport measurements alone, possible candidates include Wigner~\cite{ChuiTanatarPRL1995}, striped phases~\cite{SlutskinPRB2000} and bubble phases~\cite{FogleretalPRB1996}. Since $\rho_{2DES} \propto$ the number of topological defects $n_d$, we estimate $n_d \propto \rho_0 \exp{(A/\sqrt{n_s - n_{KT}})} \equiv N_0\exp{(A/\sqrt{n_s - n_{KT}})}$, where $\rho_0$, $A$ and $n_c$ are obtained from the fit in figure~\ref{DensityDependence}a and $N_0$ is an unknown constant. We now test the applicablity of this idea to the case of $S$, i.e., we ask whether the topological defects mediate the thermal transport as well. We note that this idea can readily be used to explain the enhancement in $|S|$ since the number of topological defects $n_d < n_s$ which, according to equation~\ref{DrudeMottformula}, would bring about such an enhancement. To check this quantitatively, we plot $S$ as a function of $n_d$ in figure~\ref{SvsNd}. We immediately find that in the region where $S$ deviates from $S_d$, oscillations notwithstanding, it obeys a $1/n_d$ envelope. This strengthens the notion that the topological defects and not the bare electrons are responsible for transport in the 2DES (see equation~\ref{DrudeMottformula}). The value of $N_0$ used in figure~\ref{SvsNd} is obtained by fitting the observed $S$ to equation~\ref{DrudeMottformula} and assuming the electronic values of $e$ and $m$. We note that this need not be the case though it is beyond the scope of the present measurements to resolve this issue. The topological defects behave essentially as non-interacting, free quasiparticles and this also explains the observed metallic $T$-dependence of $S$. 

We finally address the oscillations in $S$ as a function of $n_s$. While the position and magnitude of these oscillations are device-dependent (see reference~\cite{Narayan}) we note that in all the devices measured, they are completely absent in $\rho_{2DES}$. This is in stark contrast to Coulomb Blockade oscillations in which $\sigma$ and $S$ oscillate synchronously as the system parameters are varied (see for example refs.~\cite{ScheibnerPRL2005}). We also note that previour conductance measurements on similar 2DESs~\cite{ArindamJPhysC2004, SeigertPRB2008, BaenningerPRB2008} have shown intriguing oscillations that, once again, cannot be reconciled with Coulomb Blockade behaviour and suggest, rather, the formation of pinned charge-density-wave phases. It is possible that the oscillations reflect the entropy of the 2DES~\cite{GargNJP2011}, though further experiments are required to resolve this matter.

\section{Conclusions}

Our electrical and thermal transport measurements combine to present a self-consistent picture in which novel quasiparticles born of inter-electron interactions mediate transport. These quasiparticles impart a metallic character to the 2DES and enhance its thermopower by more than two orders of magnitude. The thermopower of the 2DES is also seen to oscillate strongly as the density is varied and these oscillations are completely absent in the resistivity. While the origin of these oscillations is unknown, they are a striking example of the thermopower being more sensitive to interactions than the electrical resistivity.

\begin{acknowledgements}

We acknowledge funding from the UK-India Education and Research Initiative (UKIERI), the Department of Science and Technology (DST), India and the Engineering and Physical Sciences Research Council (EPSRC), UK. VN acknowledges a Fellowship from the Herchel Smith Fund.

\end{acknowledgements}

\pagebreak

\end{document}